\documentclass{PoS}
\pdfoutput=1

\usepackage[utf8]{inputenc}
\usepackage{booktabs}

\title{Fragmentation functions of charged hadrons}

\ShortTitle{Fragmentation functions of charged hadrons}

\author{\speaker{Emanuele R. Nocera}\\
        Rudolf Peierls Centre for Theoretical Physics, 
        University of Oxford,\\ 
        1 Keble Road, Oxford OX1 3NP, United Kingdom\\
        E-mail: \email{emanuele.nocera@physics.ox.ac.uk}}

\abstract{I present a preliminary determination of the Fragmentation
          Functions (FFs) of unidentified charged hadrons at 
          next-to-leading order in quantum chromodynamics.
          The analysis is based on hadron production cross section data in 
          single-inclusive electron-positron annihilation.
          It extends a recent determination of the FFs of identified 
          charged pions, charged kaons and protons/antiprotons performed
          by the NNPDF Collaboration.
          I illustrate the quality of the FFs determined in this analysis and 
          show how they describe the charged hadron spectra measured in 
          proton-(anti)proton collisions.}

\FullConference{XXV International Workshop on Deep-Inelastic Scattering 
                and Related Subjects\\
		3-7 April 2017\\
		University of Birmingham, UK}

\begin{document}

In a recent paper~\cite{Bertone:2017tyb}, the NNPDF Collaboration presented
NNFF1.0, a determination of the Fragmentation Functions (FFs) of charged
pions, charged kaons and protons/antiprotons.
The analysis included a comprehensive set of cross section measurements
for single-inclusive hadron production in electron-positron annihilation (SIA). 
It was performed within the NNPDF methodology - a fitting framework designed to 
provide a statistically sound representation of FF uncertainties with a minimal 
procedural bias - at leading, next-to-leading and next-to-next-to-leading order
(LO, NLO and NNLO) in perturbative Quantum Chromodynamics (QCD).
The systematic inclusion of higher-order QCD corrections allowed the 
ensuing NNFF1.0 FF sets to describe the data very well, 
including those points at rather small hadron momentum fractions $z$.
Together with existing sets of unpolarised~\cite{Ball:2017nwa}
and polarised~\cite{Nocera:2014gqa} Parton Distribution Functions (PDFs),
FFs and PDFs are available from a common, mutually consistent framework
for the first time.

A number of hard-scattering processes, probing nucleon momentum, spin, flavour,
spatial distributions and the dynamics of nuclear 
matter~\cite{Albino:2008aa} require the knowledge of FFs. 
In some of them, charged hadrons are inclusively measured in the 
final state, without identifying the hadronic species.
In this case, although pions, kaons and protons/antiprotons dominate
the hadron yields, FFs accounting also for the residual hadron fraction 
should be used.

In this contribution, I extend the NNFF1.0 analysis to a preliminary 
determination of the FFs of unidentified charged hadrons, though only at NLO.
As in NNFF1.0, I restrict to SIA, and I consider flavour-untagged (inclusive)
and tagged measurements performed by experiments at CERN 
(ALEPH~\cite{Buskulic:1995aw}, DELPHI~\cite{Abreu:1998vq,Abreu:1997ir} 
and OPAL~\cite{Ackerstaff:1998hz,Akers:1995wt}),
DESY (TASSO~\cite{Braunschweig:1990yd}) 
and SLAC (TPC~\cite{Aihara:1988su} and SLD~\cite{Abe:2003iy}).
The included data set is summarised in Tab.~\ref{tab:datasets}, where I
specify, in the first five columns, the name of the experiment, 
the corresponding publication reference,
the measured observable, the center-of-mass energy $\sqrt{s}$, and the number 
of data points included in the fit after (before) applying kinematic cuts.
On top of total inclusive and tagged cross section measurements, 
longitudinal inclusive and tagged cross section data are also available
for unidentified charged hadrons, in contrast to identified charged pions, 
charged kaons and charged protons/antiprotons.
They are proportional respectively to the structure functions $F_2^{h^++h^-}$
and $F_L^{h^++h^-}$.
Contrary to the former, the latter includes a non-vanishing 
$\mathcal{O}(\alpha_s)$ contribution, with $\alpha_s$ the QCD strong coupling, 
already at LO~\cite{Rijken:1996vr,Mitov:2006wy}.
This helps constrain the gluon FF, especially
since the lack of precise data over a wide range of energies makes 
it only roughly determined by evolution.

%-------------------------------------------------------------------------------
\begin{table}[!t]
\renewcommand{\arraystretch}{2}
\centering
\scriptsize
\begin{tabular}{lclccc}
\toprule
Experiment & Reference & Observable & 
$\sqrt{s}$ [GeV] & $N_{\rm dat}$ & $\chi^2/N_{\rm dat}$\\
\midrule
TASSO14 & \cite{Braunschweig:1990yd}
& $\frac{1}{\sigma_{\mathrm{tot}}}\frac{d\sigma^{h^\pm}}{dz}$
& 14.00 
& 15 (20)
& 1.23\\
TASSO22 & \cite{Braunschweig:1990yd}
& $\frac{1}{\sigma_{\mathrm{tot}}}\frac{d\sigma^{h^\pm}}{dz}$
& 22.00 
& 15 (20)
& 0.51\\
TPC & \cite{Aihara:1988su}
& $\frac{1}{\sigma_{\mathrm{tot}}}\frac{d\sigma^{h^\pm}}{dz}$
& 29.00 
& 21 (34)
& 1.65\\
TASSO35 & \cite{Braunschweig:1990yd}
& $\frac{1}{\sigma_{\mathrm{tot}}}\frac{d\sigma^{h^\pm}}{dz}$
& 35.00 
& 15 (20)
& 1.14\\
TASSO44 & \cite{Braunschweig:1990yd}
& $\frac{1}{\sigma_{\mathrm{tot}}}\frac{d\sigma^{h^\pm}}{dz}$
& 44.00 
& 15 (20)
& 0.68\\
ALEPH & \cite{Buskulic:1995aw}
& $\frac{1}{\sigma_{\mathrm{tot}}}\frac{d\sigma^{h^\pm}}{dz}$
& 91.20 
& 32 (35)
& 1.04\\
& \cite{Buskulic:1995aw}
& $\frac{1}{\sigma_{\mathrm{tot}}}\frac{d\sigma^{h^\pm}_L}{dz}$
& 91.20 
& 19 (21)
& 0.36\\
DELPHI & \cite{Abreu:1998vq}
& $\frac{1}{\sigma_{\mathrm{tot}}}\frac{d\sigma^{h^\pm}}{dp_h}$
& 91.20 
& 21 (27)
& 0.65\\
& \cite{Abreu:1998vq}
& $\left.\frac{1}{\sigma_{\mathrm{tot}}}\frac{d\sigma^{h^\pm}}{dp_h}\right|_{uds}$
& 91.20 
& 21 (27)
& 0.17\\
& \cite{Abreu:1998vq}
& $\left.\frac{1}{\sigma_{\mathrm{tot}}}\frac{d\sigma^{h^\pm}}{dp_h}\right|_{b}$
& 91.20 
& 21 (27)
& 0.82\\
& \cite{Abreu:1997ir}
& $\frac{1}{\sigma_{\mathrm{tot}}}\frac{d\sigma^{h^\pm}_L}{dz}$
& 91.20 
& 20 (22)
& 0.72\\
& \cite{Abreu:1997ir}
& $\left.\frac{1}{\sigma_{\mathrm{tot}}}\frac{d\sigma^{h^\pm}_L}{dz}\right|_{b}$
& 91.20 
& 20 (22)
& 0.44\\
OPAL & \cite{Ackerstaff:1998hz}
& $\frac{1}{\sigma_{\mathrm{tot}}}\frac{d\sigma^{h^\pm}}{dz}$
& 91.20 
& 20 (22)
& 2.41\\
& \cite{Ackerstaff:1998hz}
& $\left.\frac{1}{\sigma_{\mathrm{tot}}}\frac{d\sigma^{h^\pm}}{dz}\right|_{uds}$
& 91.20 
& 20 (22)
& 0.90\\
& \cite{Ackerstaff:1998hz}
& $\left.\frac{1}{\sigma_{\mathrm{tot}}}\frac{d\sigma^{h^\pm}}{dz}\right|_{c}$
& 91.20 
& 20 (22)
& 0.61\\
& \cite{Ackerstaff:1998hz}
& $\left.\frac{1}{\sigma_{\mathrm{tot}}}\frac{d\sigma^{h^\pm}}{dz}\right|_{b}$
& 91.20 
& 20 (22)
& 0.21\\
& \cite{Akers:1995wt}
& $\frac{1}{\sigma_{\mathrm{tot}}}\frac{d\sigma^{h^\pm}_L}{dz}$
& 91.20 
& 20 (22)
& 0.31\\
SLD & \cite{Abe:2003iy}
& $\frac{1}{\sigma_{\mathrm{tot}}}\frac{d\sigma^{h^\pm}}{dp_h}$
& 91.28 
& 34 (40)
& 0.75\\
& \cite{Abe:2003iy}
& $\left.\frac{1}{\sigma_{\mathrm{tot}}}\frac{d\sigma^{h^\pm}}{dz}\right|_{uds}$
& 91.28 
& 34 (40)
& 1.03\\
& \cite{Abe:2003iy}
& $\left.\frac{1}{\sigma_{\mathrm{tot}}}\frac{d\sigma^{h^\pm}}{dz}\right|_{c}$
& 91.28 
& 34 (40)
& 0.62\\
& \cite{Abe:2003iy}
& $\left.\frac{1}{\sigma_{\mathrm{tot}}}\frac{d\sigma^{h^\pm}}{dz}\right|_{b}$
& 91.28 
& 34 (40)
& 0.97\\
\midrule
Total dataset & & & & 471 (527) & 0.83\\
\bottomrule
\end{tabular}
\caption{\small The data set included in this analysis of FFs for unidentified 
 charged hadrons.
 For each experiment, I indicate the publication reference, the measured 
 observable, the center-of-mass energy $\sqrt{s}$, the number of data points 
 included after (before) kinematic cuts, and the $\chi^2$ per number of data 
 points, $\chi^2/N_{\rm dat}$.}
\label{tab:datasets}
\end{table}
%-------------------------------------------------------------------------------

The theoretical and methodological details of this analysis, including the
description of the experimental observables in terms of the FFs,
their evolution through {\tt APFEL}~\cite{Bertone:2013vaa},
their parametrisation, and the fitting procedure, are exactly the same
as in the NNFF1.0 analysis.
Kinematic cuts are choosen consistently.
Data points below $z_{\rm min}$, with $z_{\rm min}=0.02$ for experiments 
at $\sqrt{s}=M_Z$, $z_{\rm min}=0.075$ for all other experiments, and 
$z_{\rm max}=0.9$ are not included in this fit, as in NNFF1.0.

\begin{figure}[!t]
\centering
\includegraphics[scale=0.135,angle=270,clip=true,trim=2cm 0 4cm 0]{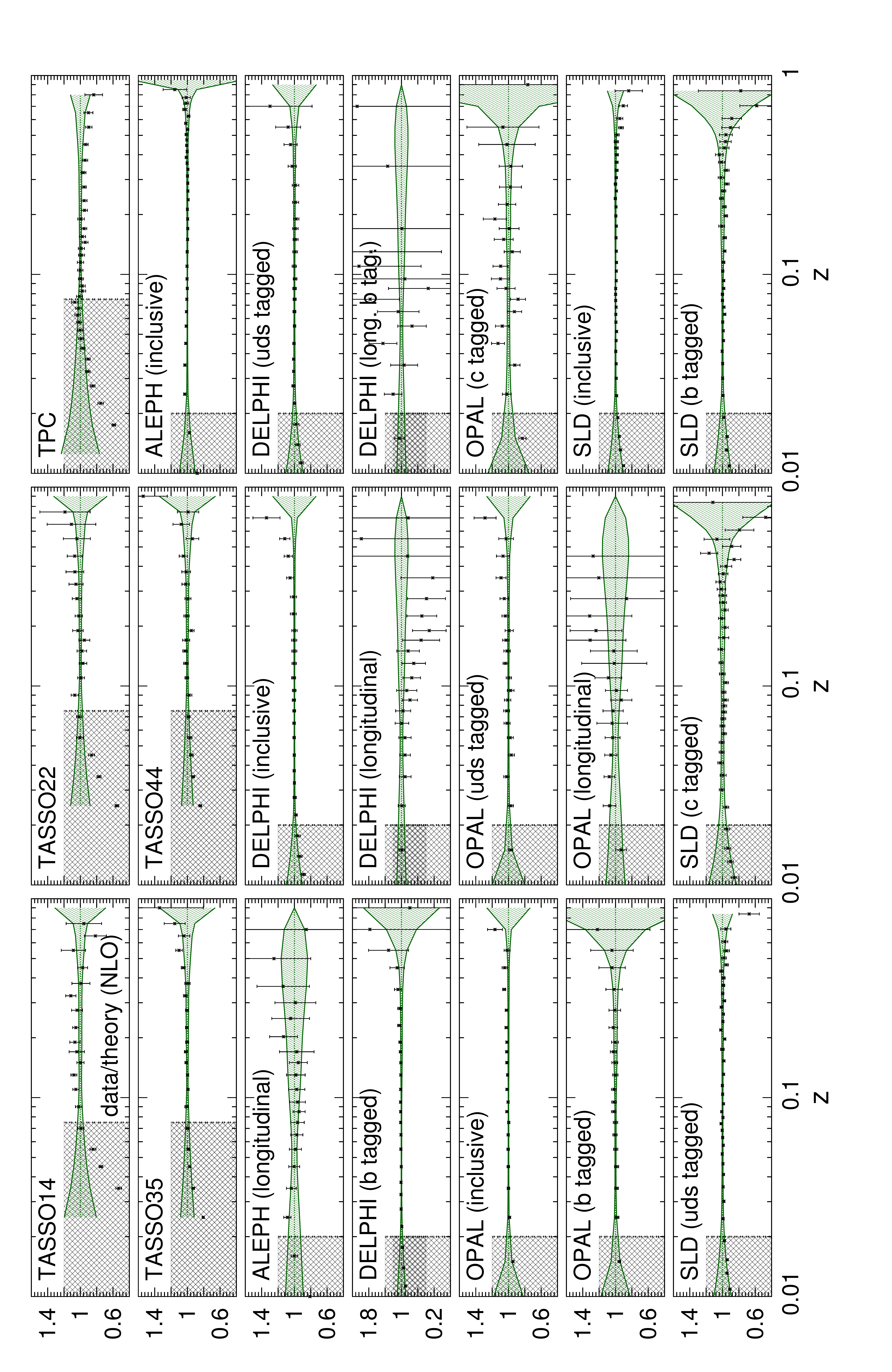}\\
\caption{The data/theory ratio for the unidentified light charged hadron 
data included in this analysis. Theoretical predicitons are computed at NLO
with the corresponding best-fit FFs. Bands correspond to one-$\sigma$ FF 
uncertainties. Shaded areas indicate the regions excluded by kinematic cuts.}
\label{fig:datatheory}
\end{figure}

In the last column of Tab.~\ref{tab:datasets}, I report the $\chi^2$ per
number of data points, $\chi^2/N_{\rm dat}$, for both the individual and 
total data sets.
The data/theory ratio is diplayed for each data set in 
Fig.~\ref{fig:datatheory}.
Bands correspond to one-$\sigma$ FF uncertainties, and shaded areas indicate 
the regions excluded by kinematic cuts.
As is apparent from both Tab.~\ref{tab:datasets} and Fig.~\ref{fig:datatheory},
the overall fit quality, measured by the total $\chi^2/N_{\rm dat}=0.83$,
is very good.
All data sets are very well described, with the exception of the 
OPAL inclusive data set.
The large value of the $\chi^2$ per number of data points, 
$\chi^2/N_{\rm dat}=2.41$, is determined by a handful of rather precise 
data points at large values of $z$.
A similar behaviour was already observed in the 
NNFF1.0 analysis for the individual hadron species, where a certain amount
of tension between the OPAL data and the other data sets at $\sqrt{s}=M_Z$
was found.
The data points below $z_{\rm min}$ are reasonably described by the extrapolated 
FFs for all data sets at $\sqrt{s}=M_Z$.

%-------------------------------------------------------------------------------
%-------------------------------------------------------------------------------

\begin{figure}[!p]
\centering
\includegraphics[scale=0.46,angle=270,clip=true,trim=0 0 1.2cm 0]{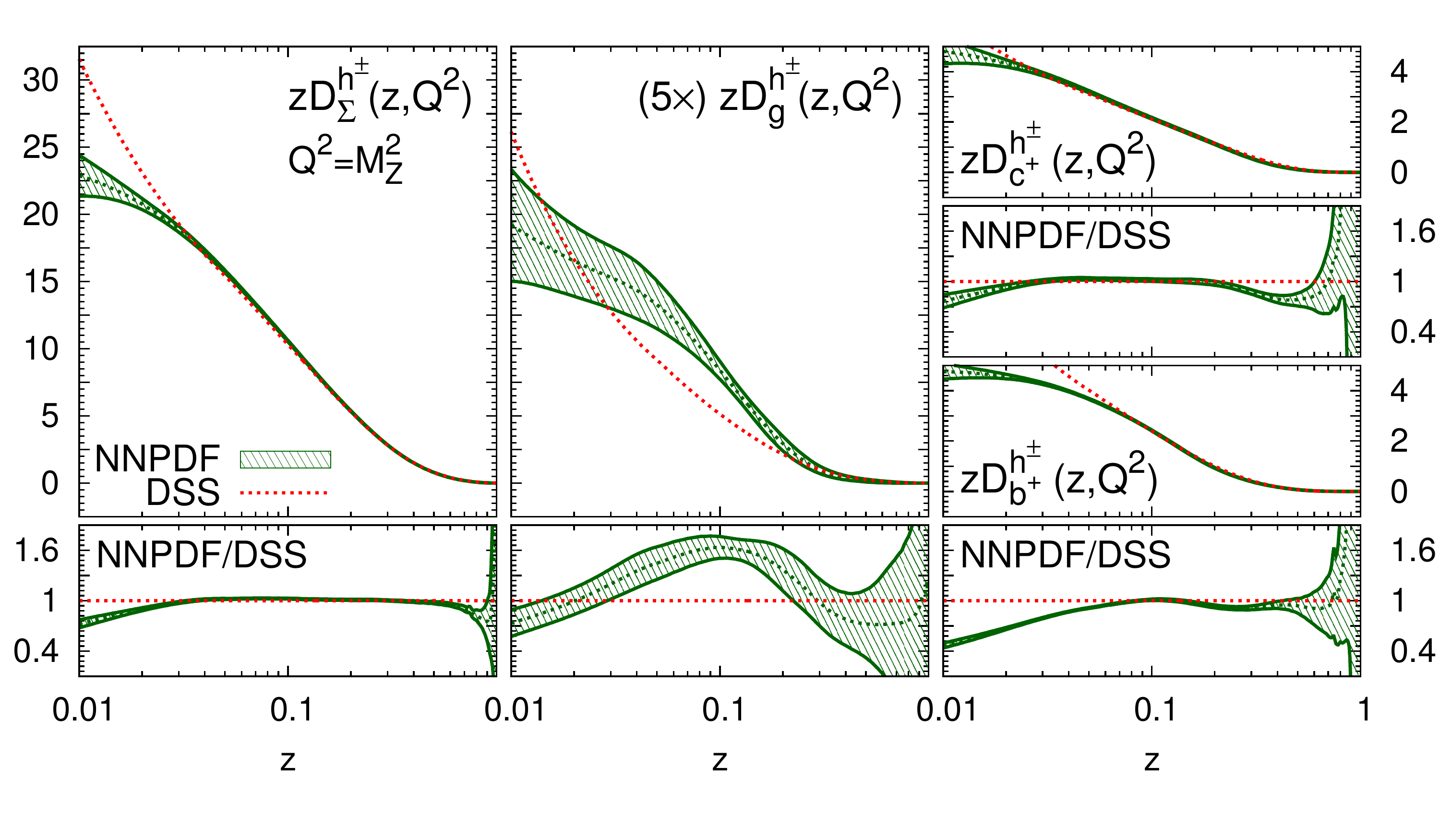}\\
\caption{Comparison between the singlet, gluon, and total charm and bottom
FFs for the unidentified charged hadrons obtained from this analysis 
(denoted as NNPDF) and the DSS set at $Q^2=M_Z^2$; 
NNPDF uncertainties correspond to one-$\sigma$ bands.
The ratio to the DSS central value is also displayed below each FF.}
\label{fig:HAFFs}
\end{figure}

The FFs for the singlet, $D_{\Sigma}^{h^\pm}=\sum_q^{n_f}D_q^{h^\pm}+D_{\bar{q}}^{h^\pm}$ 
(with $n_f$ the number of active flavours), gluon, $D_g$, and total charm and 
bottom, $D_{q^+}^{h^\pm}=D_{q}^{h^\pm}+D_{\bar{q}}^{h^\pm}$ (with $q=c,b$) are 
displayed in Fig.~\ref{fig:HAFFs} at $Q^2=M_Z^2$ as a function of $z$.
The results from this analysis (denoted as NNPDF) include the one-$\sigma$
uncertainty bands.
They are compared to the central value from the DSS 
analysis~\cite{deFlorian:2007ekg}, to which they 
are also normalised in each panel below the corresponding FFs.
Uncertainty bands on the DSS results are not shown because they are not 
available to the author.

As is apparent from Fig.~\ref{fig:HAFFs}, the most relevant discrepancy
between this analysis and DSS is in the gluon FF, which is harder in the 
former than in the latter at small-to-medium values of $z$, 
$0.03\lesssim z\lesssim 0.3$.
It is instead softer at large values of $z$, $0.3\lesssim z\lesssim 0.9$, 
although compatible with the DSS central value within uncertainties.
A similar behaviour was observed for the gluon FF of charged pions
in NNFF1.0~\cite{Bertone:2017tyb}.
At small-$z$, it was recognised to be a consequence of the less conservative 
small-$z$ kinematic cut adopted in~\cite{Bertone:2017tyb} 
with respect to~\cite{deFlorian:2007ekg}; at large-$z$ of the more flexible 
NNPDF parametrisation.
Furthermore, this determination, despite being based on a wider set
of SIA data than~\cite{deFlorian:2007ekg}, does not include proton-(anti)proton
collider data. 
This data is directly sensitive to the gluon FF, it was taken into account 
in~\cite{deFlorian:2007ekg}, and may also account for the differences observed.
All other FFs in Fig.~\ref{fig:HAFFs} are in reasonable agreement with the DSS
central value, except at small $z$, again because of the different 
kinematic cut adopted in the two analyses.

%-------------------------------------------------------------------------------
%-------------------------------------------------------------------------------
%-------------------------------------------------------------------------------
\begin{figure}[!p]
\centering
\includegraphics[scale=0.14,angle=270]{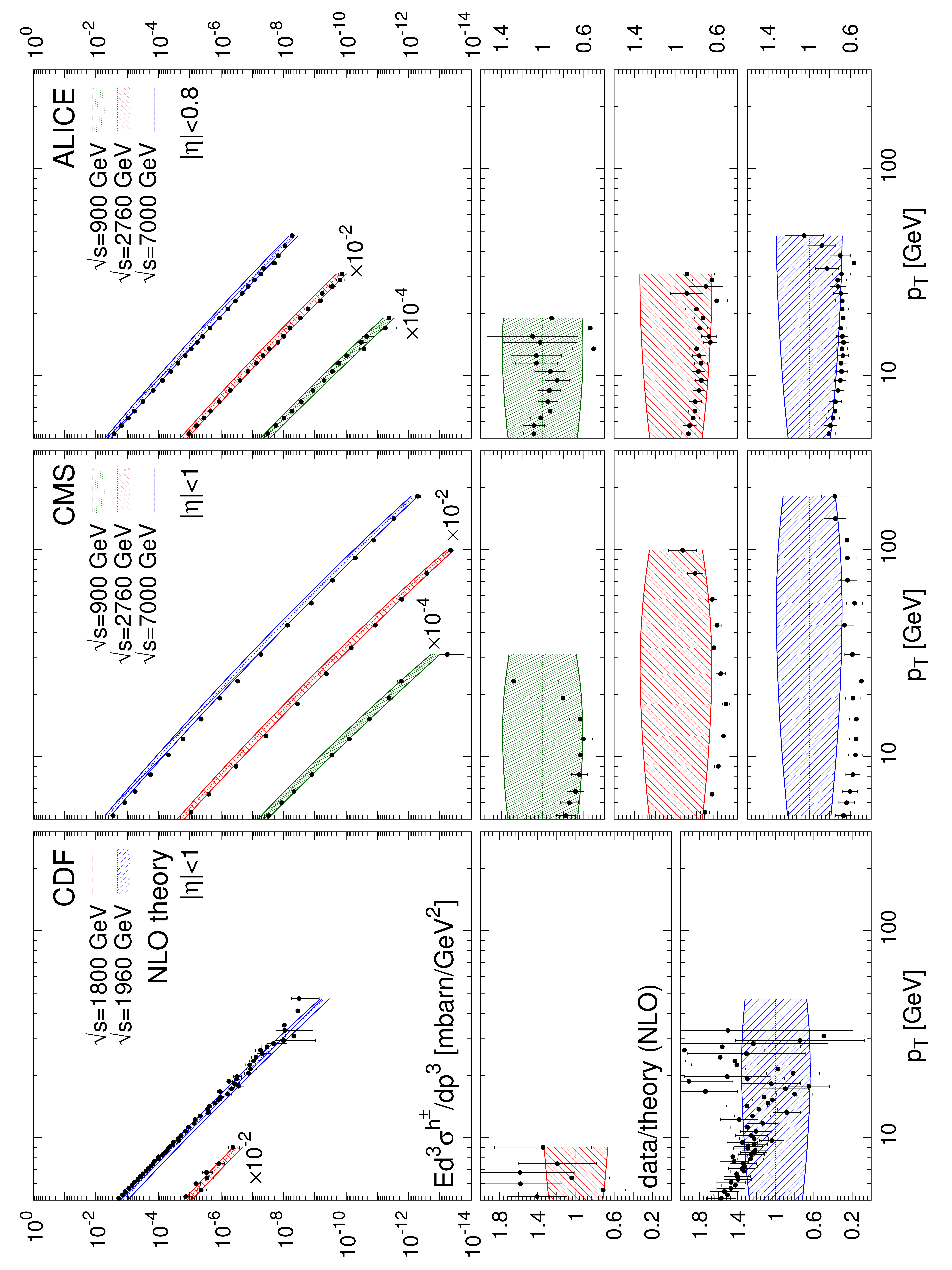}\\
\caption{Differential cross section data for the inclusive charged hadron 
spectra measured by CDF, CMS and ALICE in proton-(anti)proton collisions at
different center-of-mass energies $\sqrt{s}$ integrated over the rapidity 
$\eta$.
The data is compared to the NLO predictions obtained with the FFs determiend 
in this analysis.
The corresponding data/theory ratio is also shown.
One-$\sigma$ bands include the FF uncertainty only.}
\label{fig:ppdata}
\end{figure}
%-------------------------------------------------------------------------------

As an example of a possible application, the FFs determined in this analysis
are used to compute predictions for the differential cross section spectra 
as a function of the transverse momentum of the hadron, $p_T$.
The results are obtained at NLO with the code presented in~\cite{Jager:2002xm}
and the PDF central value from the NNPDF3.1 set~\cite{Ball:2017nwa}.
They are compared to measurements performed in proton-antiproton collisions
by the CDF experiment at $\sqrt{s}=$1.8 and 1.96 TeV~\cite{Abe:1988yu}, 
and in proton-proton collisions by the CMS~\cite{Chatrchyan:2011av} 
and ALICE~\cite{Abelev:2013ala} experiments at $\sqrt{s}=$0.9, 2.76 and 7 TeV.
The data/theory ratio is also shown.
One-$\sigma$ uncertainty bands include the FF uncertainty only.

The data-theory agreement displayed in Fig.~\ref{fig:ppdata} is reasonable
in almost all cases.
Theoretical predictions tend to overshoot experimental data at the highest
center-of-mass energies.
However they remain mutually compatible within uncertainties, except in the 
case of the CMS experiment at $\sqrt{s}=$2.76 and 7 TeV.
A similar behaviour was pointed out in~\cite{dEnterria:2013sgr}, where 
all the FF sets available at that time were shown to grossly overshoot
charged hadron spectra cross section data.
In this analysis such an effect, though still partly present, is mitigated.
The reason is related to the gluon FF, to which the observable in 
Fig.~\ref{fig:ppdata} is primarily sensitive.
The gluon FF determined in this analysis is indeed rather different from 
the DSS one (and from the one in other similar analyses), 
see Fig.~\ref{fig:HAFFs}.
Furthermore, it is affected by larger uncertainties, as explicitly shown for
identified hadrons in~\cite{Bertone:2017tyb}.
Finally, the uncertainty of the data, especially from CMS and ALICE,
is rather narrower than the uncertainty due to the FF.
This suggests that the data will carry in a large amount of 
information on the gluon FF once included in a global FF fit~\cite{Bertonep}. 

I thank V.~Bertone and L.~Rottoli for their collaboration 
in sorting out the data and predictions shown in Fig.~\ref{fig:ppdata}.
This work is supported by the UK STFC Rutherford Grant ST/M003787/1.

\end{document}